\documentstyle[twocolumn,prb,aps,psfig]{revtex}
\begin{document}
 \twocolumn[\hsize\textwidth\columnwidth\hsize\csname @twocolumnfalse\endcsname
\draft
\begin{title}
{Dephasing of Electrons by Two-Level Defects in Quantum Dots}
\end{title}
\author{ Kang-Hun Ahn}
\address{ Max-Planck-Institut f\"ur Physik Komplexer Systeme, 
N\"othnitzer Str. 38, 01187 Dresden, Germany      \\ }
\author{ Pritiraj Mohanty }
\address{ Condensed Matter Physics 114-36, 
California Institute of Technology, Pasadena, CA 91125 \\ }
\date{\today}
\maketitle
  
\widetext

\begin{abstract}
The electron dephasing time $\tau_{\phi}$ in a diffusive quantum dot
is calculated by considering the interaction between the electron and
dynamical defects, modelled as two-level system. Using  the standard 
tunneling model of glasses, we obtain a linear temperature dependence 
of $1/\tau_{\phi}$, consistent with the experimental observation.
However, we find that, in order to obtain dephasing times on the
order of nanoseconds, the number of two-level defects needs to be 
substantially larger than the typical concentration in glasses. We also 
find a finite system-size dependence of $\tau_{\phi}$, which can 
be used to probe the effectiveness of surface-aggregated defects. 
\vskip 0.1in

\vskip 0.1in
\end{abstract}
]
\narrowtext

\newpage
\section{Introduction}

Interference of the electron's paths in a mesoscopic conductor results 
in various quantum phenomena such as the universal conductance 
fluctuation, persistent current, and weak localization. 
In all these phenomena, the 
dephasing time $\tau_{\phi}$ appears as a typical time scale over 
which the electronic trajectories 
have interference; weak localization correction to 
conductivity---for example, is conventionally used for the experimental 
determination of the dephasing time \cite{altshuler}.
In the moderate temperature range, experimentally-determined\cite{wind} 
values of $\tau_{\phi}$ in diffusive metals are found to be 
in excellent agreement 
with the theoretical predictions of $\tau_\phi$ due to electron-electron 
interaction \cite{altshuler}. While it is theoretically expected that 
$\tau_\phi \rightarrow \infty$ as $T\rightarrow 0$ in the absence 
of other external sources of dephasing, $\tau_\phi$ is found to saturate 
at low temperatures in {\it almost} all experiments\cite{bergmann}, including the 
recent carefully-performed experiments\cite{mohanty1,lin,doug,saclay}. 
This severe discrepancy between theory and experimental observation 
of low temperature saturation 
has fast become a topic of controversy \cite{golubev,aleiner} 
surrounding the question whether the idea\cite{mohanty1} and the 
theory\cite{golubev} of zero-point fluctuations of the electromagnetic 
field created by the electron-electron interaction as a source of 
dephasing are tenable on general grounds. This poses 
a serious problem as zero temperature dephasing of electrons has been 
argued to be relevant to the problems of persistent current in normal 
metals \cite{persistent-current}, the low temperature metal/insulator, 
quantum-Hall/insulator and superconductor/insulator 
transitions \cite{mohanty2,kapitulnik,meir}, 
and transport through various normal-metal/superconductor hybrid junctions
\cite{mohanty3,ova}; but the 
most unsettling consequence is the negation of the fundamental 
premise upon which the theories---and hence our understanding---of 
metals and insulators are based: the many-body Fermi liquid picture.  

Among various sustained efforts to find a zero temperature dephasing 
mechanism other than electron-electron interaction, 
dephasing due to dynamical defects inside the conductor has been 
recently argued \cite{imry,zawadowski} to be important to the 
saturation problem. Low-energy excitations of the dynamical defects 
are usually modelled by two-level systems (TLS). Invoked some
three decades ago, first by Anderson, Halperin and Varma \cite{anderson}, 
and also by Phillips \cite{phillips0}, the tunneling model of TLS has been quite 
successful in explaining various anomalies in the acoustic, dielectric and
thermodynamic properties of structural glasses and other 
amorphous solids \cite{phillips}. 

Imry, Fukuyama and Schwab \cite{imry} have recently
suggested that the saturation behavior may have the same origin as 
the $1/f$ conductance noise, arising from the two-level defects. 
Zawadowski, von Delft and Ralph \cite{zawadowski} have argued  
that the apparent saturation of $\tau_{\phi}$ may be caused by the 
two-channel Kondo effect due to electron-TLS scatterings.  
However, it was pointed out\cite{mohanty2,lin} that hysteresis or 
switching behavior, expected from the effects of TLS, was not observed 
in experiments. In addition, various concentration-dependent 
Kondo-like bulk trends anticipated in these theories were also not 
observed in the experiments.

In this paper, we investigate the role of two-level defects in the dephasing
of electrons in {\it quantum dots}. In the recent experiments from the 
Marcus group, Huibers and coworkers have observed the saturation of  
dephasing time in open  quantum dots\cite{huibers} below 0.1 K 
along with a strong temperature dependence above 0.1 K.
In addition to this experiment, saturation of $\tau_\phi$ in quantum dots 
has also been reported in other experiments\cite{bird,{clarke},{buks}}, 
although the physical meaning of $\tau_{\phi}$ in these set-ups are not 
the same. If two-level defects are responsible for the saturation of 
dephasing time in this experiments\cite{huibers},
 it is natural to suppose that just above 0.1 K  the linear temperature
dependence should be explained by two-level defects as well.
Our calculations indeed show that the dephasing rate due to the TLS
does have a linear temperature dependence.
However, we find that the magnitude of the dephasing by two-level defects
is too small to explain the experimentally observed dephasing time of
nanoseconds. This implies that either other mechanisms are more effective 
or surface-aggregated two-level defects play a dominant role; defects on 
a disordered surface are likely to have unusual distributions in their 
splitting energies; we suggest that the surface defects can 
be experimentally probed by measuring the size dependence of 
dephasing time. 


\begin{figure}
\centerline{\psfig{figure=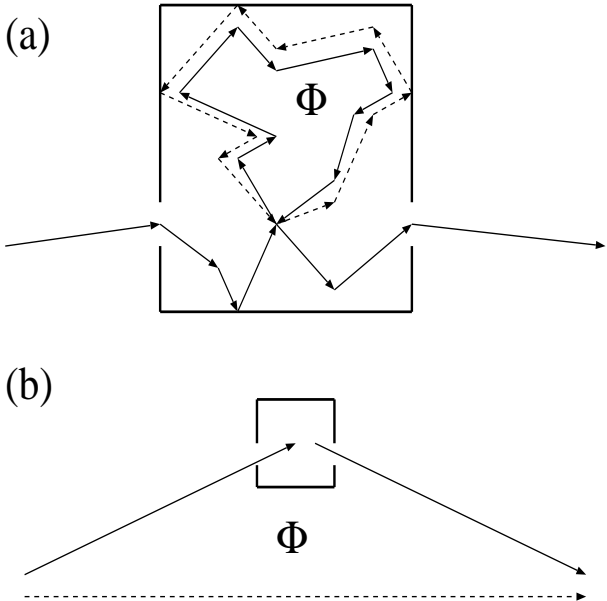,height=8cm,width=0.45\textwidth}}
\caption{Two types of electron interferometry involving quantum dots (a) Both paths
are inside the dot. (b) One of the interferring path is outside the dot.}
\end{figure}

Consideration of two-level defects in quantum dots for their
dominant role in dephasing is motivated by the experimentally observed 
tell-tale signs  of TLS in quantum dots: hysteresis and switching 
behaviors, which have been in fact detected in various quantum-dot experiments 
\cite{zimmerli,{furlan},{grupp}} unlike in the experiments on higher dimensional 
diffusive metals. In a quantum dot, usually the Thouless 
energy $E_{T}$ is the largest energy scale unlike the diffusive metallic 
case\cite{altshuler,golubev}. Therefore, the results obtained 
in the diffusive metallic cases \cite{altshuler,golubev} 
cannot be applied (even after the appropriate dimensional considerations) 
to quantum dots. 
Thouless energy is  defined by $E_{T}\equiv \hbar D/L^{2}$, where
$D$ is the diffusion constant and $L$ is the typical system size.

Dephasing generally describes the loss of coherence or 
suppression of interference. Accordingly,  it is important to know 
which kind of paths are considered before defining
the typical time scale of the loss of interference along these paths.
In Fig.~1, we show two kinds of 
electron interferometers involving quantum dots.
In this paper, we are concerned with pairs of time-reversed 
 paths which return to the origin in a diffusive system (Fig. 1. (a)).
These time-reversed paths enclose magnetic flux; their interference
manifests in the weak localization correction to conductivity.
These paths are chosen, for the problem at hand, because their 
contribution to conductivity does not vanish even after disorder averaging.
In the interferometry discussed in this paper, 
change of the mean conductance at a finite field from its zero field value,
$\delta g=\big<\big< g \big>\big>_{B\neq 0}-\big< \big<g \big>\big>_{B=0}$,
 can be used to
extract the dephasing time from experimental data, 
 where $\big<\big<...\big>\big>$ means disorder averaging.
Using the phenomelogical random matrix theory (RMT) in 
Refs.\cite{baranger,brouwer}, $\tau_{\phi}$
can be defined, for instance, by the formula
\begin{eqnarray}
\delta g \approx \frac{e^{2}}{h}
\left( \frac{N}{2N+\frac{2\pi\hbar}{\tau_{\phi}\Delta}} \right),
\label{gcahnge}
\end{eqnarray}
where $N$ is the number of channels connected to a quantum dot.
Although, the formulae for the conductance change
are model dependent \cite{baranger,brouwer},
the difference in the equations in these models
are not significant to the interpretation of $\tau_{\phi}$ measured 
in experiments \cite{huibers}.
In this work, we will refer to $\tau_{\phi}$ which were obtained
through $\delta g$ as in Ref.\cite{huibers} without discussing 
how $\delta g$ is related to $\tau_{\phi}$ any further.

Our calculation of dephasing time is similar to the calculation of
dephasing time by Stern, Aharonov and Imry \cite{stern}.
Based on the interference of two time-reversed trajectories,
we calculate a typical time scale over which the environmental 
state remains in the initial state.
Dephasing rate $1/\tau_{\phi}$ due to electron-electron interactions 
in diffusive quantum dots 
has been calculated by Sivan and coworkers\cite{sivan2}:
\begin{eqnarray}
 \frac{1}{\tau_{\phi}}\Big|_{e-e} 
\sim \delta_{1}
\left(\frac{\large{\epsilon}}{E_{T}} \right)^{2}, \quad\quad \epsilon >> k_{B}T; 
\label{eqsivan}
\end{eqnarray}
\noindent
where  $\epsilon$ is the excitation energy
of the particle and $\delta_{1}$ is the mean level spacing.
However,  it should be noted that the direct application of 
Eq.(\ref{eqsivan}) to experimental data \cite{huibers} is difficult,
because  it is not meaningful to
estimate the temperature dependence of $\tau_{\phi}$
by merely replacing $\epsilon$ with $\sim  k_{B}T$ in 
Eq.(\ref{eqsivan})\cite{comment1}.

While the interference paths considered here
are inside the dot as depicted in Fig.1.(a), there are 
experiments\cite{buks,yacoby} 
associated with other types of
interference involving one path through the dot and another 
outside the dot (Fig. 1. (b)).
Interference appears as a periodic Aharonov-Bohm
oscillation rather than the fluctuations 
in the electric current as in the case of Fig.1. (a).
In experiments,  dephasing manifests in the non-ideality
( $< $ 1) of the visibility of the interference pattern\cite{buks}.
Theoretically, dephasing rate involved in the geometry shown in
Fig. 1. (b) is calculated as the level broadening
\cite{levinson,aleiner3,buttiker}.
In these considerations\cite{levinson,aleiner3,buttiker}, level broadening
is due to the gauge fluctuations of a 0D-quantum state where
the electron dynamics inside the dot is neglected.
There are other ways to define dephasing time.  
In the work of Sivan and coworkers\cite{sivan1},  dephasing time 
is considered to be the quasiparticle life time, measured in the tunneling experiments.
In the work of Bird and coworkers\cite{bird}, conductance fluctuation 
was used to measure $\tau_{\phi}$ at very strong magnetic fields.
Nevertheless, we will refrain from discussing these different views 
of dephasing time in this paper.

The organization of the paper is as follows. In section II, we describe 
how dephasing time is calculated in a general framework.
In section III, interaction between the two-level defects and
electrons is discussed. In section IV, the return probability is calculated 
in order to obtain $\tau_{\phi}$.
In section V, we discuss dephasing by two-level defects with 
widely distributed  energies.
In section VI, dephasing by
two-level defects with narrowly distributed energies are discussed.

\section{ Interference and dephasing of particle's trajectory }

Let us consider the event that the electron is at the position ${\bf r}$ at 
an initial time $t=0$, and it
arrives at the position ${\bf r^{\prime}}$ by diffusive motion after 
a time $\tau_{0}$.
The environmental state changes from $\eta$ to $\eta^{\prime}$ in
this process; the corresponding probability amplitude of the event is
$\rho( {\bf r^{\prime}}, {\bf r},\eta^{\prime},\eta;\tau_{0})$. 

The description of the suppression of interference in electron's paths  
by the electron-TLS interaction can be considered in two 
different approaches: 

\begin{enumerate}

\item[(i)] The electron in the two different paths  
produces two different time-dependent electric fields on TLS, thereby
TLS go to different states, which suppresses interference.

\item[(ii)]  The fluctuating dipole moment of TLS  produces the time-dependent
electric field, thereby the electron in the two different paths
gains random phase, which also suppresses interference.

\end{enumerate}

In general, these two approaches are not equivalent,
because the presence of the electron induces a back reaction from the
TLS environment.
However, in the presence of weak interaction between the particle and
the environment, it is known that either the two description are equivalent,
or at least they give the same dephasing rate up to the second order in 
the interaction\cite{stern2}. In this work, we use approach (i).
Following the scheme of Chakravarty and Schmid\cite{chakravarty}, we 
use semiclassical approximation on particle's trajectory and 
we consider quantum mechanical evolution of the TLS (environment) states.
We further assume that the TLS environment does not influence the classical paths
of the electron, therefore the diffusive electron motion comes from only static disorder.
Under certain conditions, the two-level defects might be able to effectively change
the semiclassical paths of the electrons. In that case, one may estimate $\tau_{\phi}$ 
by calculating electron-TLS inelastic scattering time. However, 
$\tau_{\phi}$ begins to lose its meaning as a dephasing time, 
since we lose the semiclassical picture of the electron's path.


The tunneling motion in the TLS environment is to be
described in a fully quantum mechanical way.
To this end, we consider the time-dependent potential 
$ \hat{V} ({\bf {r}}(t))$ 
exerted by the moving electron of the path ${\bf r}(t)$
on a two-level defect.
The probability amplitude is given by
\begin{eqnarray}
\rho( {\bf r^{\prime}}, {\bf r},\eta^{\prime},\eta;\tau_{0})
=\sum_{j}A_{j}({\bf r^{\prime}},{\bf r};\tau_{0})e^{i S_{j}}
\big< \eta^{\prime} | 
U_{j}(\tau_{0})
| \eta  \big>,
\label{p_amp}
\end{eqnarray}
where $A_{j}$ and $S_{j}$
  are the corresponding
amplitude and  action of a classical electron's 
trajectory labeled by $j$.
$U_{j}(\tau_{0})$ is a time-evolution operator 
(in the interaction picture)
of the environmental  state
associated with the electron trajectory ${\bf r}_{j}(t)$
\begin{eqnarray}
U_{j}(\tau_{0})={\rm \hat{T}} \exp\left[{\frac{i}{\hbar}
\int_{0}^{\tau_{0}} V_{I} ({\bf r}_{j}(t),t) dt}\right],
\end{eqnarray}
where ${\rm \hat{T}}$ is the time-ordering operator and 
\begin{eqnarray}
 \hat{V}_{I}({\bf r}_{j}(t),t) 
=e^{\frac{i}{\hbar}H_{env}t} \hat{V} ({\bf r}_{j}(t)) 
e^{-\frac{i}{\hbar}H_{env}t}.
\end{eqnarray}

The probability $P( {\bf r^{\prime}}, {\bf r},\eta;\tau_{0})$
of finding the particle at ${\bf r^{\prime}}$ after time 
$\tau_{0}$, initially at ${\bf r}$ with the environment
in the initial state $|\eta\big>$, is
given by the sum of the absolute square of the probability amplitudes 
over the final states of the environment;
\begin{eqnarray}
P_{\eta}( {\bf r^{\prime}}, {\bf r};\tau_{0})
&=&\int d\eta^{\prime}
|\rho( {\bf r^{\prime}}, {\bf r},\eta^{\prime},\eta;\tau_{0})|^{2}
\\ \nonumber &=&
\sum_{j}|A_{j}({\bf r^{\prime}},{\bf r};\tau_{0})|^{2} 
\\
&+&\sum_{j\neq k} A_{j}A_{k}^{*}e^{i(S_{j}-S_{k})}
\big< \eta \big | 
U_{k}^{\dagger}(\tau_{0})
U_{j}(\tau_{0}) \big| \eta \big>,
\label{p_amplitude}
\end{eqnarray}
using the completeness relation for the environmental states.

The return probability $P_{{\bf r},\eta}(\tau_{0})$ 
of an electron is defined by the probability
of finding the electron at position ${\bf r}$ 
 after time $\tau_{0}$ 
which was initially at the same position
with the environmental state $\big| \eta\big>$;
\begin{eqnarray} 
P_{{\bf r},\eta}(\tau_{0})
&=& 
P_{\eta}({\bf r^{\prime}}, {\bf r};\tau_{0})
 \big|_{  {\bf r^{\prime}}  \rightarrow {\bf r} } 
\\ &=&
\sum_{j}|A_{j}({\bf r},{\bf r};\tau_{0})|^{2}
\label{return_1}
\\ \nonumber
&+&\sum_{j}|A_{j}|^{2}\big<\eta\big|        
\frac{
U_{j^{T}}^{\dagger} U_{j}+
U_{j}^{\dagger} U_{j^{T}}}{2}
 \big|\eta\big>+\cdots,
\end{eqnarray}
where $j^{T}$ denotes the time-reversed path of $j$.
The first term in Eq.(\ref{return_1}) is termed as the classical return
probability $P^{class}_{\bf r}(\tau_{0})$. 
The second term comes from the interference of the pair of time-reversed 
paths. The remaining terms that do not appear in the above
equation vanish upon ensemble averaging over disorder due to the
random differences in its classical action $S_{j}-S_{k}$ for
 $ k\neq j,j^{T}$. The coherent part 
$P^{coh}_{{\bf r},\eta}(\tau_{0})$
of the return probability is the second term
in Eq.(\ref{return_1});
\begin{eqnarray}
P^{coh}_{{\bf r},\eta}(\tau_{0})
= \sum_{j}|A_{j}({\bf r},{\bf r};\tau_{0})|^{2}{\rm Re}\big<\eta\big|        
U_{j}^{\dagger}(\tau_{0}) U_{j^{T}}(\tau_{0})  \big|\eta\big>.
\label{return_2}
\end{eqnarray}

\noindent
Now, the dephasing time can be defined as a time scale for 
$P^{coh}(\tau_{0})$
to vanish  with a decreasing function such as 
$\exp{(-\tau_{0}/\tau_{\phi})}$.
But for the present purpose, the particular exponential form of 
time dependence is not needed.

\section{Electron-TLS interactions}

We consider two-level tunneling systems (TLS)\cite{phillips}
 as the environment for an electron in the quantum dot.
Let us first consider TLSs which have asymmetry energy $\Delta$ and
the tunnel splitting energy $\Delta_{0}$.
The TLSs are assumed to be randomly distributed over the dot with 
their electric dipole moments  randomly oriented.
We will assume the dipole moment is not too strong so that we do not
have to consider interaction among the TLS.
The density of the TLS will be assumed to be not too high so that 
multiple scattering events between the electron and the dipoles can be
neglected.
Within this approximation, we calculate the return  probability
of an electron 
in the presence of a single TLS, thereby we extend the reults to the 
case of many randomly distributed TLSs.
The Hamiltonian of the TLS 
can be written 
in terms of the localized wave functions of
the double well potential and also in terms of the eigenenergy basis
\begin{eqnarray}
H_{TLS}&=&\frac{1}{2} \left(\begin{array}{rr} \Delta &  \Delta_{0} \\
 \Delta_{0}   &  -\Delta \end{array} \right) 
\rightarrow
\frac{1}{2} \left(\begin{array}{rr} E &  0 \\
 0   &  -E \end{array} \right) 
\label{htls}
\end{eqnarray}
where
$E=\sqrt{\Delta^{2}+\Delta_{0}^{2}}$ and the transformation denoted by the arrow 
means localized wavefunction representation $\rightarrow $
 eigen wavefunction representation.
The dipole strength operator $\hat{p}$ is defined in the eigen wavefunction representation:
\begin{eqnarray}
\hat{p} = p_{0}\left(\begin{array}{rr} 1 &  0 \\
 0   &  -1 \end{array} \right)
 \rightarrow
p_{0} \left(\begin{array}{rr} \Delta/E &  \Delta_{0}/E \\
 \Delta_{0}/E   &  -\Delta/E \end{array} \right) 
\label{poperator}
\end{eqnarray}
where $p_{0}$ is the dipole moment when the particle is located
in one of the wells of the defect potential.
In the following sections, we will use the eigen wavefunction representation
in which $H_{TLS}$ is diagonal.
The TLS Hamiltonian will be used 
for the environment Hamiltonian $H_{env}=H_{TLS}$.

TLS dipole at the position ${\bf R}$ feels the electric field
${\bf  E( R ) }$ produced by
 the moving electron. The resulting interaction energy can be
expressed by the operator  $\hat{V}({\bf r}_{j}(t))$;
\begin{eqnarray}
\hat{V}({\bf r}_{j}(t)) 
=-\hat{p}{\rm \bf n}\cdot {\bf  E}({\bf R})
=-\hat{p}{\rm \bf n}\cdot {\bf \nabla}_{\bf R}
V_{c}(|{\bf R}-{\bf r}_{j}(t)|),
\label{vrt}
\end{eqnarray}
where $\hat{p}$ is the dipole moment operator for the TLS, 
which is along the direction of unit vector ${\bf n}$.
$V_{c}$ is Coulomb interaction potential;
\begin{eqnarray}
V_{c}({\bf R}-{\bf r}_{j}(t))
=\frac{e}{\epsilon^{*}|{\bf R}-{\bf r}_{j}(t)| }
\approx \frac{1}{L^{d}}
\sum_{\bf q}v_{\bf q}e^{i{\bf q}\cdot ({\bf r}_{j}-{\bf R})},
\label{vc}
\end{eqnarray} 
where $e$ is the electric charge, $\epsilon^{*}$ is the  dielectric constant
of the dot material, $L$ is the linear system size, and 
$d$ is the spatial dimensionality of the dot ($d$=2, 3).
Here, $v_{\bf q}$ is the finite Fourier transform
of  Coulomb potential;
\begin{eqnarray}
v_{\bf q}&=&\int_{dot} d{\bf r} e^{-i{\bf q \cdot r}}
\frac{e}{\epsilon^{*} |{\bf r}|}
\\
&=&\frac{2\pi e}{\epsilon^{*}q}~~~~(2D,~~q \neq 0)
\label{vq2}
\\
&=&\frac{4\pi e}{\epsilon^{*}q^{2}}~~~~(3D,~~q\neq 0)
\label{vq3}
\end{eqnarray}
We use the discrete values of ${\bf q}=\frac{2\pi}{L}(m,n,k)$ for
the three-dimensional case, and
 ${\bf q}=\frac{2\pi}{L}(m,n)$ for the two-dimensional case where
$l,m$, and $n$  are integers.
By inserting Eq.(\ref{vc}) into Eq.(\ref{vrt}), we get 
\begin{eqnarray}
\hat{V}({\bf r}_{j}(t)) 
=\frac{1}{L^{d}}
{\hat p} \sum_{\bf q} \big(i {\rm \bf n}\cdot {\bf q}~ v_{\bf q}
e^{-i{\bf q}\cdot{\bf R}} \big)
 e^{i{\bf q} \cdot {\bf r}_{j}(t)}.
\label{vrt_det}
\end{eqnarray}

Derivation of the Eq.(\ref{vrt_det}) is based on the semiclassical approximation
of the electron's motion and the unscreened dipole moment of TLS.
From a pure quantum-mechanical point of view,
one can consider the TLS-electron interaction similar to its treatment 
in metalic glasses\cite{black}. 
When the TLS 
has two positions
${\bf R^{\pm}}
={\bf R}\pm {\bf d}/2$, the purely quantum mechanical TLS-electron
interactions $\hat{V}_{QM}$ are written as\cite{black} 
\begin{eqnarray}
\nonumber
\hat{V}_{QM}&=&\frac{1}{L^{d}}\frac{\hat{p}}{p_{0}}
\sum_{q}2i\mu_{q}
e^{-i{\bf q}\cdot{\bf R} }
\sin({\bf q}\cdot{\bf d}/2) 
\\ 
&\times&  \sum_{k}c_{k}^{\dagger}c_{k+q},
\label{hprime}
\end{eqnarray}
where $c^{\dagger}_{k}$ ($c_{k}$) is the electron creation(annihilation) operator 
with momentum $k$, and $\mu_q$ is the Fourier transform
of the ionic potential.
However, the pure quantum-mechanical approach is not reliable, because
the concept of dephasing  becomes ambiguous as we leave the semiclassical 
approximation, which has been pointed out in Ref.\cite{cohen}. 
Here, we merely get some useful informations by comparing the quantum mechanical
Hamiltonian in Eq.(\ref{hprime}) to our semiclassical potential
in Eq.(\ref{vrt_det}).
Since ${\bf q}\cdot {\bf d} << 1$ in quantum dots
(this is true--generally speaking, 
when the Fermi wavelength is much larger than d), 
$\sin({\bf q}\cdot{\bf d}/2)
\approx {\bf q}\cdot{\bf d}/2
={\bf n}\cdot {\bf q}\frac{p_{0}}{2e}$.
Eq.(\ref{hprime}) can be understood as 
the interaction between the TLS-dipole  coupled to an effective
electric field produced by the electron.

One of the two differences between Eq.(\ref{vrt_det}) and Eq.(\ref{hprime})
is that the electron interacts with the ion through a screened
interaction $\mu_{q}$ in Eq.(\ref{hprime}) rather
than the direct Coulomb interaction $v_{q}$ as in Eq.(\ref{vrt_det}).
The specific form of $\mu_{q}$ is not known though it is
 expected to be less than $v_{q}$.
If  a screened interaction $\mu_{q}$ is used, then the calculated dephasing rate 
$1/\tau_{\phi}$ would be smaller
than that with the unscreened interaction $v_{q}$.
In this work, we will use $v_{q}$ instead of $\mu_{q}$.
The second difference, which is rather important, 
is the quantum mechanical nature of the electron motion in Eq.(\ref{hprime}).
$\hat{V}({\bf r}_{j}(t))$ in Eq.(\ref{vrt_det}) is understood 
as $\big< \psi_{j}({\bf r},t)\big|\hat{V}_{QM}\big|
 \psi_{j}({\bf r},t)\big>$, where 
 $\psi_{j}({\bf r},t)$ is the time-dependent wavefunction which describes
a wave-packet corresponding to the trajectory $j$;
\begin{eqnarray}
 e^{i{\bf q} \cdot {\bf r}_{j}(t)}
\approx 
\big< \psi_{j}({\bf r},t)\big|
  \sum_{k}c_{k}^{\dagger}c_{k+q}\big|
 \psi_{j}({\bf r},t)\big>. 
\end{eqnarray}
At a finite temperature $T$, 
the wave-packet state
 $|\psi_{j}({\bf r},t)\big> $  will consist of 
mostly the eigenenergy states 
with energies limited within $\epsilon_{F}\pm k_{B}T$.
Therefore, the time dependence of 
$ e^{i{\bf q} \cdot {\bf r}_{j}(t)}$
will be limited by the frequency  windows $|\omega| < k_{B}T$.
Though we use the semiclassical approach in 
Eq.(\ref{vrt_det}),
the frequency cut-off will be performed in our semiclassical 
calculation later. 

\section{The Return Probability}

For the simplicity of calculation, 
we suppose that the TLS is initially in the state $|->$, 
an eigenstate of the Hamiltonian in Eq.(\ref{htls}) with the
eigenenergy $E_{-}$
(the case with $|+>$ can be calculated in a similar way).
The time evolution operator $U_{j}(t)$
($U_{j}(t)\big|-\big>=c_{+}(t)\big|+\big>+c_{-}(t)\big|-\big>$)
 of the TLS corresponding to
the electron path $j$ can be written as  
\begin{eqnarray}
\left( 
\begin{array}{r} 
c_{+}(t) \\ c_{-}(t)
\end{array}
\right)={\rm \hat{T}}
\exp
\left[-\frac{i}{\hbar} \int_{0}^{t}dt^{\prime}
{\bf V_{I}(r_{j}(t))}
\right]
\left( 
\begin{array}{r} 
0 \\ 1
\end{array}
\right), 
\end{eqnarray}
where 
\begin{eqnarray}
\nonumber
&&{\bf V_{I}(r_{j}(t))}= \\
&&\left(
\begin{array}{rr}  
\big<+\big|V_{I}(r_{j}(t^{\prime}),t^{\prime})\big|+\big> &
\big<+\big|V_{I}(r_{j}(t^{\prime}),t^{\prime})\big|-\big>
\\
\big<-\big|V_{I}(r_{j}(t^{\prime}),t^{\prime})\big|+\big> &
\big<-\big|V_{I}(r_{j}(t^{\prime}),t^{\prime})\big|-\big>
\end{array}
\right)
\end{eqnarray}

\noindent
To find the corresponding 
time evolution of the TLS states for the time-reversed paths,
($U_{j^{T}}(t)\big|-\big>=d_{+}(t)\big|+\big>+d_{-}(t)\big|-\big>$),
one obtains a similar form by using
${\bf r}_{j^{T}}(t) ={\bf r}_{j}(\tau_{0}-t)$.
Expanding 
$\big< -\big| U_{j}^{\dagger}(\tau_{0}) U_{j^{T}}(\tau_{0})
\big| -\big> = c_{+}^{*}(\tau_{0})d_{+}(\tau_{0}) 
+ c_{-}^{*}(\tau_{0})d_{-}(\tau_{0})$ up to the second
order in interaction $\hat{V}$, we get
\begin{eqnarray}
\nonumber
&&{\rm Re} 
\big< -\big| U_{j}^{\dagger}(\tau_{0}) U_{j^{T}}(\tau_{0}) \big| -\big> 
\\ \nonumber
&=&1+
\\
\nonumber
&&
\frac{1}{\hbar^{2}}
\int_{0}^{\tau_{0}} dt
\int_{0}^{\tau_{0}} dt^{\prime}
\left[
\cos(\Omega(t+t^{\prime}-\tau_{0})) 
-\cos(\Omega(t-t^{\prime})) \right] \\ 
&&\times \big<-\big|\hat{V}({\bf r}_{j}(t))\big|+\big>
\big<+\big|\hat{V}({\bf r}_{j}(t^{\prime}))\big|-\big>.
\label{repcoh}
\end{eqnarray}
where
$\hbar\Omega=E=E_{+}-E_{-}=\sqrt{\Delta^{2}+\Delta_{0}^{2}}$,
and we used the relation
$\hat{V}({\bf r}_{j}(t))^{\dagger} =\hat{V}({\bf r}_{j}(t))$ in the 
last equality.

Now from Eq.(\ref{vrt_det}),  Eq.(\ref{repcoh}) and 
Eq.(\ref{return_2}), one can get
the coherent return probability in 
\begin{eqnarray}
\nonumber
&&P^{coh}_{{\bf r}} (\tau_{0})
=\sum_{j}|A_{j}({\bf r},{\bf r};\tau_{0})|^{2}
+\frac{\big|\big<+\big|\hat{p}\big|-\big>\big|^{2}}
{ 3 \hbar^{2}L^{2d} }
\int_{0}^{\tau_{0}} dt
\int_{0}^{\tau_{0}} dt^{\prime}
\\ \nonumber
&&
\left[
\cos(\Omega(t+t^{\prime}-\tau_{0})) 
-\cos(\Omega(t-t^{\prime})) \right] 
\\ &&\times
\sum_{\bf q}q^{2}\big| v_{q}\big|^{2}
\sum_{j}|A_{j}({\bf r},{\bf r};\tau_{0})|^{2}
e^{i{\bf q}\cdot({\bf r}_{j}(t)-{\bf r}_{j}(t^{\prime}))}.
\label{pcohr}
\end{eqnarray}

Here, we omitted the subscript $\eta$ in $P^{coh}_{{\bf r},\eta}$ 
for the TLS state, because both of the initial states of the TLS
 $\big| \eta \big> =\big| \pm \big> $  give rise to the same 
expression.
$P^{coh}_{\bf r} (\tau_{0})$ in Eq.(\ref{pcohr}) is the value
averaged over the TLS position.
 Here, we used the disorder average over TLS
$<<\exp[i{\bf R}\cdot({\bf q}+{\bf q^{\prime}})]>>
=\delta_{\bf q,-q^{\prime}}$.
The factor 3 in Eq.(\ref{pcohr}) comes from the average over the
orientation of the TLS dipoles.

In the case of $\sqrt{D\tau_{0}} >$ {\it l} 
({\it l} is the mean free path), 
the sum over the classical paths  which
 appears in Eq.(\ref{pcohr}) can be written as a path integral
using the 
Wiener measure\cite{feynman,schulman,chakravarty}.
The path integral can be calculated as
\begin{eqnarray}
\nonumber
&&\sum_{j}|A_{j}({\bf r},{\bf r};\tau_{0})|^{2}
e^{i{\bf q}\cdot({\bf r}_{j}(t)-{\bf r}_{j}(t^{\prime}))}
\\
\label{wien}
&=&\int_{{\bf x}(0)={\bf r}}
^{{\bf x}(\tau_{0})={\bf r}}
 D[{\bf x}(\tau)]
\\ \nonumber
&&\exp(-\frac{1}{4D}\int_{0}^{\tau_{0}} d\tau |\dot{\bf x}(\tau)|^{2} ) 
\exp(i {\bf q} \cdot ({\bf x}(t)-{\bf x}(t^{\prime}))  )
\\ \nonumber
 &=&
\frac{1}{\mathcal V}\sum_{| {\bf q}^{\prime}| < \pi/{\it l}  }
\exp(-D|{\bf q^{\prime}}|^{2}\tau_{0})
e^{-D(|{\bf q}|^{2}-2{\bf q}\cdot{\bf q^{\prime}})
|t-t^{\prime}|},
\end{eqnarray}
where we used the boundary conditions of a quantum dot by a 
rectangular box of the volume ${\mathcal V}=$
$L\times L\times$ ($L$ or $a << L$ for 2D dots).
Several remarks are in order.
Eqs. (\ref{wien}) are valid only in diffusive regimes.
The summation over the discretized momentum variables
${\bf q}$ and ${\bf q}^{\prime}$ is understood to be limited
by $\pi/{\it l}$.
The contributions from ballistic regime, which are supposed to be
small when $\sqrt{D \tau_{\phi}} >> {\it l} $, are neglected in this work.
The last equality in the Eqs. (\ref{wien}) is obtained for ${\bf r}=0$.
To simplify the calculation, from here on, let us consider the return
probability of the particle at the origin ${\bf r}=0$.

Then, the classical return probability 
$P^{class}(\tau_{0})$ for ${\bf r}=0$ 
is given by
\begin{eqnarray}
\nonumber
P^{class}(\tau_{0})&=&
\sum_{j}|A_{j}({\bf r},{\bf r};\tau_{0})|^{2}\Big|_{{\bf r}=0}
\\ 
&=&\frac{1}{\mathcal V}\sum_{|\bf q| < \pi/{\it l}}
\exp (-D|{\bf q}|^{2}\tau_{0} ).
\end{eqnarray}

Inserting Eq.(\ref{wien}) into Eq.(\ref{pcohr}), 
we get the coherent part of the return probability
\begin{eqnarray}
\nonumber
P^{coh}(\tau_{0})
&=&\frac{1}{\mathcal V}\sum_{\bf q^{\prime}}
\exp(-D|{\bf q^{\prime}}|^{2}\tau_{0})
\times
\\
&& \Big[
1+ \sum_{\bf q}
\frac{\big| \big< +\big| \hat{p}\big|- \big> \big|^{2}q^{2}v_{q}^{2} }
{3\hbar^{2} L^{2d}}
\int_{0}^{\tau_{0}}dt^{+}\int_{-t^{+}}^{t^{+}}dt^{-} \\ 
\nonumber
&&\big[
\cos\Omega(t^{+}-\tau_{0}) -
\cos\Omega t^{-} \big]
e^{
-D(|{\bf q}|^{2}-2{\bf q}\cdot{\bf q^{\prime}})
|t^{-}|} \Big],
\label{pcohtau0}
\end{eqnarray}
where we used the change of variables of $t^{+}=t+t^{\prime}$ and
$t^{-}=t-t^{\prime}$. 

We restrict to the ergodic regime $\tau_{0} >\tau_{D}$, therefore
significant contribution comes only from ${\bf q^{\prime}}=0$.
Then, we obtain
\begin{eqnarray}
\nonumber
P^{coh}(\tau_{0})
&=&\frac{1}{\mathcal V}
\Big[
1+ \sum_{\bf q}
\frac{\big| \big< +\big| \hat{p}\big|- \big> \big|^{2}q^{2}v_{q}^{2} }
{3\hbar^{2} L^{2d}}
\int_{0}^{\tau_{0}}dt^{+}\int_{-t^{+}}^{t^{+}}dt^{-}
\\ \nonumber
&&\big[
\cos\Omega(t^{+}-\tau_{0}) -
\cos\Omega t^{-} \big]
\\&&
 \frac{1}{2\pi}
\int d\omega
\frac{\exp(i\omega |t^{-}|)}{i\omega +D q^{2}}
 \Big].
\label{pcohtau0_2}
\end{eqnarray}

The frequency of the time-dependent electric field produced by the 
electron is not infinitely large but has a cut-off. 
By assuming the electron to be in equilibrium with other electrons at
temperature $T$, the high frequency cut-off of $\omega$ is given by
$k_{B}T$ ($|\omega| <k_{B}T $); this is true at temperatures 
that are not too low.
Note that because of the finite size of the system, $qv_{q}=0$ for $q=0$.
Therefore, there is no divergence at low frequencies and 
the low-frequency cut-off of $\omega$ does not play an important role.
By integrating Eq.(\ref{pcohtau0_2}) 
with the condition  of 
$|\omega| <k_{B}T $, 
we obtain
the coherent part of the return probability $P^{coh}$, which decays as
 $\propto 1-\tau_{0}/\tau_{\phi}+\cdots$. We have now defined 
$\tau_{\phi}$ as the dephasing time. 
The  dephasing rate $1/\tau_{\phi}$ 
from a randomly distributed 
TLS with an asymmetry energy $\Delta$ and a tunnel splitting 
$\Delta_{0}$ is given by
\begin{eqnarray}
\frac{1}{\tau_{\phi}(\Delta,\Delta_{0})}
&=&\frac{2
\big| \big< +\big| \hat{p}\big|- \big> \big|^{2}}{3\hbar^{2}L^{2d}}
\sum_{\bf q}q^{2}
v_{q}^{2}\frac{Dq^{2}}{\Omega^{2}+(Dq^{2})^{2}}
\label{deph1}
\\
&\approx&
\frac{2p_{0}^{2}
\Delta_{0}^{2}}{3\hbar^{2}D(\Delta^{2}+\Delta_{0}^{2})}
\sum_{0< |\bf q| <\pi/{\it l}}v_{q}^{2}L^{-2d}.
\label{deph2}
\end{eqnarray} 
We used $\Omega \tau_{D} << 1$ in the second equality in 
Eq.(\ref{deph2}).  
$1/\tau_{\phi}$  obtained above is valid only when $k_{B}T > \hbar \Omega$,
while in the other case $P^{coh}(\tau_{0})$ is an oscillating function
of $\tau_{0}$ with a small amplitude. Therefore
$1/\tau_{\phi}$ ($ k_{B}T < \hbar \Omega $) is negligible.

By inserting Eq.(\ref{vq2}) and Eq.(\ref{vq3}) into Eq.(\ref{deph2})
when $k_{B}T > \sqrt{\Delta^{2}+\Delta_{0}^{2}}$ 
we get
\begin{eqnarray}
\frac{1}{\tau_{\phi}(\Delta,\Delta_{0})}
&=&
\frac{2p_{0}^{2}e^{2}
\Xi_{d}
\Delta_{0}^{2}}{3\hbar^{2}D\epsilon^{*2}L^{2}(\Delta^{2}+\Delta_{0}^{2})}
\label{deph3}
\end{eqnarray}
where $d=2, 3$ is the spatial dimension of the quantum dot, and
\begin{eqnarray}
\Xi_{2}&=&\sum_{0< m^{2}+n^{2} <(L/l)^{2}}
\frac{1}{m^{2}+n^{2}}
\\
\Xi_{3}&=&\frac{1}{\pi^{2}}\sum_{0< m^{2}+n^{2}+k^{2} <(L/l)^{2}}
\frac{1}{m^{2}+n^{2}+k^{2}}.
\end{eqnarray}

\section {Dephasing by two-level defects with widely distributed energies }

We can generalize $1/\tau_{\phi}$ to
the case where the TLSs are distributed with a distribution function
$f(\Delta,\Delta_{0})$,  
\begin{eqnarray}
\frac{1}{\tau_{\phi}}=
{\mathcal V} 
\int d\Delta \int d\Delta_{0} \frac{1}{\tau_{\phi}(\Delta,\Delta_{0})}
f(\Delta,\Delta_{0}).
\end{eqnarray}
\noindent
By inserting Eq.(\ref{deph3}) into the above equation, we get

\begin{eqnarray}
\frac{1}{\tau_{\phi}}=\frac{L^{d-2}}{D}
\frac{2p_{0}^{2}e^{2} \Xi_{d} a^{3-d}}
{3\hbar^{2}\epsilon^{*2}} S(T),
\label{deph_main}
\end{eqnarray}
where
\begin{eqnarray}
\nonumber
S(T)=\int d\Delta &&\int d\Delta_{0} \frac{\Delta_{0}^{2}}
{\Delta^{2}+\Delta_{0}^{2} }
\\ &&
f(\Delta,\Delta_{0})
\theta( k_{B}T -\sqrt{\Delta^{2}+\Delta_{0}^{2}}),
\label{ST}
\end{eqnarray}
where $a$ is the thickness of the dot in case $d=2$.
It is interesting to note 
that at $d=2$, the dephasing time does not depend on the dot area.

To calculate $\tau_{\phi}$, we use the standard tunneling model
for the two-level defects \cite{anderson,phillips0}.
The essential postulate in this theory is the uniform distribution of
the tunneling parameter $\lambda$ associated with the tunnel splitting
$\Delta_{0} \propto e^{-\lambda}$.
The energy distribution function $f(\Delta,\Delta_{0})$ in this case
is written as
\begin{eqnarray}
f(\Delta,\Delta_{0})=\frac{\bar{P}}{\Delta_{0}}.
\end{eqnarray}
Furthermore, it is also assumed that $\Delta_{0}$ has a nonzero 
minimum value $\Delta_{0,min}$.
By applying this distribution,
we find;
\begin{eqnarray}
S(T)
&=&
\bar{P}\int^{k_{B}T}_{\Delta_{0,min}} d\Delta_{0}
\int^{\sqrt{(k_{B}T)^{2}-\Delta_{0}^{2} } } 
_{-\sqrt{(k_{B}T)^{2}-\Delta_{0}^{2}}} 
d\Delta \frac{\Delta_{0}}{\Delta^{2}+\Delta_{0}^{2}}
\\
&=&2\bar{P}\Delta_{0,min}{\mathcal F}(k_{B}T/\Delta_{0,min}),
\end{eqnarray}
where
\begin{eqnarray}
{\mathcal F}(z)=\int_{1}^{z}dx{\rm tan}^{-1}
\frac{\sqrt{z^{2}-x^{2}}}{x}
\end{eqnarray}
Here, the above expression is valid for $k_{B}T < \Delta_{0,max} $, which
is an realistic and common assumption for the temperature below 1K.  
Note that ${\mathcal F}(z) \sim z\ln z$ when $z >> 1$, therefore in the case of
$k_{B} T  >> \Delta_{0,min} $, we expect the following  temperature dependence
\begin{eqnarray}
\frac{1}{\tau_{\phi}} \propto T \ln(\frac{k_{B}T}{\Delta_{0,min}}) ,
\end{eqnarray}
which is closer to $ \sim T$ rather than $ \sim T^{2}$.

Now let us estimate $\tau_{\phi}$ quantitatively.
We consider the experiments by Huibers and coworkers\cite{huibers}
on two-dimensional ballistic semiconductor quantum dots.
The quantum dots in the experiments\cite{huibers}
are in the ballistic regime, while our $\tau_{\phi}$ is for  diffusive 
quantum dots.
However, since the dephasing time is in the ergodic regime
($\tau_{\phi}> \tau_{D}$), the results for  diffusive
dots should be applicable to the chaotic quantum dots in the ballistic regime.
The diffusion coefficient  is obtained through the ergodic time scale
and $D\sim (E_{Th}/\delta_{1})(\hbar/2m^{*})$, 
where $m^{*}$ is the effective mass of the
electron ( for GaAs, $m^{*}=0.067m_{e}$ ), and 
$E_{Th}/\delta_{1} \sim 30$ for Ref.\cite{huibers}.
For ballistic dots, the Thouless energy is given by $\hbar v_{F}/L$.
For GaAs, $\hbar^{2}\epsilon^{*}/m^{*}e^{2} \sim 10 nm$.
A reasonable size of the dipole moment is $p_{0}\sim e \times 10^{-10}m$.
The thickness of the two-dimensional quantum dot is roughly $a\sim 10nm $.
By putting together $\Xi_{2}$ and $\ln(k_{B}T/\Delta_{0,min})$, which are
roughly $\sim 1-10$, into Eq.(\ref{deph_main}), 
we find
\begin{equation}
1/\tau_{\phi} \sim (10^{-16}-10^{-15}) m^{3}s^{-1}\bar{P}k_{B}T.
\end{equation}
\noindent
In order to obtain $\tau_{\phi}\sim 1ns$ near $T=0.1 K$, the average
concentration should be
$\bar{P}\sim (10^{48}-10^{49}) {\rm J}^{-1}m^{-3}$.
Although this number is not completely unreasonable, it is
too large to be expected from well-textured semiconductors 
used in the experiments\cite{huibers}.
For comparision, we note that glassy materials possess a 
typical TLS concentration of 
$\bar{P} \sim 10^{45}-10^{46}{\rm J}^{-1}m^{-3}$.

One may anticipate a different temperature dependence which might show the 
saturation of $\tau_{\phi}$  by
 considering the dissipative two-level system due to TLS-phonon interactions
or incoherent two-level systems due to TLS-TLS interactions. 
However, it is very difficult to expect that 
the dephasing rate is enhanced  by several orders of magnitude
by such interactions.

The large magnitude of $\bar{P}$ may be possible if 
a large enough number of two-level defects aggregate on the surface
of the quantum dots. 
This possibility can be experimentally checked by varying the system size
and the dimensionality. Using our results, 
\begin{equation}
1/\tau_{\phi}\propto L^{d-2}a^{3-d}.
\end{equation}
\noindent
For example, for a 2D quantum dot, the dephasing rate $1/\tau_{\phi}$ 
by  ``intrinsic" two-level defects will increase as the thickness $a$ of
the dot increases, whereas it will decrease with $a$ for surface defects.

\section{Dephasing by two-level defects with a narrow energy distribution}

Low-energy excitations exist in semiconductor crystals due to the tunneling of
impurity ions between equivalent interstitial lattice sites. Due to the crystal 
fields, definite positions are preferred and a wide distribution of excitation 
energies is not expected; in glasses, the wide distribution arises 
because of structural disorder. However, defects on the surface may result in 
a wider distribution of energies because of surface roughness. 
A single tunnel-splitting energy implies a narrow distribution of relaxation 
times such that the standard tunneling model, applicable to structural glasses, 
as discussed in the previous section, is not valid \cite{wurger}.

In this section, we consider a well defined tunnel-splitting energy $\Delta_0$ 
rather than a wide distribution. The asymmetry may be uniformly distributed 
with a gaussian width $\Delta_1$, usually determined from the experimental data. 
The distribution function is defined as
\begin{equation}
f(\Delta)=n_{TLS}\frac{1}{\Delta_{1}\sqrt{\pi}}
e^{-\Delta^{2}/\Delta_{1}^{2}}.
\end{equation}
\noindent
$n_{TLS}$ is the TLS density.

The function $S(T)$ defined in Eq.(\ref{ST}) in the expression for 
the dephasing rate $1/\tau_\phi$ 
is simplified to
\begin{eqnarray}
S(T)=\int
 d\Delta \frac{\Delta_{0}^{2}}{\Delta^{2}+\Delta_{0}^{2}}
f(\Delta)\theta (k_{B}T-\sqrt{\Delta^{2}+\Delta_{0}^{2}}).
\end{eqnarray}
Note that the variable $\Delta_0$ is not integrated over, in contrast to
the case for the standard tunneling model; and the final result depends
on $\Delta_0$.  
Evaluation of the above integral  yields 
\begin{eqnarray}
S(T)&\sim &n_{TLS} ~~(k_{B}T >> \Delta_{0} >> \Delta_{1} )\label{irrelevant}
\\
&\sim &\frac{\Delta_{0}}{\Delta_{1}}
n_{TLS} ~~(k_{B}T >> \Delta_{1} >> \Delta_{0} ). \label{relevant}
\end{eqnarray} 
If temperature is larger than the energy scales of TLS, then it is
possible to obtain saturation or temperature-independent 
dephasing rate $1/\tau_\phi$.
In realistic systems, $\Delta_0$ is usually a small fraction of
$\Delta_1$; thus the experimentally relevant limit is the second
case, $\Delta_1 \gg \Delta_0$, in expression (\ref{relevant}). 
The dephasing rate can now be obtained:
\begin{eqnarray}
{1 \over \tau_{\phi}} \sim (10^{-16}-10^{-15})n_{TLS}
 {\Delta_{0} \over \Delta_{1}}m^{3}s^{-1}.
\end{eqnarray}
For $\tau_{\phi}$ to be on the order of 1 ns, the two-level 
defect density should be 
\begin{equation}
n_{TLS} \sim  \frac{\Delta_{1}}{\Delta_{0}} (10^{24}-10^{25}) m^{-3}.
\label{density-single}
\end{equation}

Now let us estimate $n_{TLS}$ for a typical
single-crystal system. Single crystal silicon structures have been studied 
in this context in the temperature range of interest, below 1 K down to 5 mK. 
Both acoustic dissipation and heat capacity measurements on silicon resonators 
by Kleiman, Agnolet and Bishop \cite{kleiman} (see the corresponding estimates by 
Phillips \cite{phillips_si} and Keyes \cite{keyes}) find that the TLS density,
$n_{TLS} \sim 10^{23} m^{-3}$, with an estimated value of 
$\Delta_{1}/\Delta_{0} \sim 100$.
Now using the same value for $\Delta_{1}/\Delta_{0}$ in the expression 
(\ref{density-single}), the order-of-magnitude estimate of TLS density is
found to be $n_{TLS}\sim 10^{26}-10^{27} m^{-3}$. The required density 
needs to be at least three orders of magnitude higher than the typical 
concentration in the silicon
structures to result in a TLS-induced dephasing time $\tau_\phi \sim 1 ~ns$.
This is an unreasonably large number, even for the typical 
intentionally-doped semiconducting structures of silicon \cite{parpia}. 
Though, experimental studies of acoustic and thermal properties of gallium arsenide structures/heterostructures 
for the effects of two-level systems have not been done, in the temperature
range of interest for dephasing \cite{huibers,mohanty1}, recent studies 
on semi-insulating gallium arsenide resonators \cite{darrell} suggest that 
the typical TLS density is comparable to that in silicon.

\section{Conclusion }

We have calculated the dephasing time by assuming the presence of two-level defects
inside diffusive quantum dots.
The temperature dependence of $1/\tau_{\phi}$ is found
to be  roughly $\sim T$ for widely distributed two-level defects in the
standard tunneling model.
We find that to explain
 the size of the experimentally-observed dephasing times, we need  
a large number of two-level defects. This number is substantially 
larger than that found in glassy materials (almost by three orders of magnitude).
Therefore, it is hard to believe that
the electron dephasing is dominated by the intrinsic two-level defects at 
low-temperatures.
We have also calculated $\tau_{\phi}$ from a distribution of 
narrow  energy two-level defects, and we find a regime of temperature
independent $\tau_{\phi}$.
However, the required number of two-level defects is too large
as in the case of widely distributed TLS.
The system size dependence obtained in our calculation can be 
used to check the possibility of surface defects which are probably 
effective. Because of the large surface-to-volume ratio in 
quantum dots, it may be reasonable to assume that most of the 
defects are surface-aggregated. It will be interesting to estimate 
$\bar{P}$ or $n_{TLS}$ required for the observed low-temperature
charge noises of quantum dots and compare to the values from dephasing
time.
Unfortunately, we are not aware of any quantitative 
theory for the quantum-dot charge noise arising from the two-level defects.

\acknowledgements
 We are grateful to P. Fulde, Yu. V. Nazarov, A. D. Zaikin,
A. Stern, S. Kettemann, and S. Hunklinger for useful discussions.

{\it Note added.-} 
In a recent paper,  Aleiner, Altshuler, and Galperin\cite{galperin} 
have analyzed the relevance of TLS for electron dephasing.
Although they use a different approach
and evaluate $\tau_\phi$ for different systems  
(metals not quantum dots), their conclusions are similar to ours---that is,  
a substantially large concentration of TLS, ${\bar P}$,  
much larger than the typical values in metallic glasses 
is required for the quantitative explanation of the saturation 
observed in experiments on metallic wires\cite{mohanty1} 
by two-level systems\cite{imry}.


\begin{references}

\bibitem{altshuler} B. L. Altshuler, A. G. Aronov, and 
D. E. Khmelnitskii, Solid State Commun. {\bf 39},
619 (1981);J. Phys. C {\bf 15}, 7367 (1982);
B. L. Altshuler and A. G. Aronov, in 
{\em Electron-electron interactions in disordered systems}, eds.
A. L. Efros and M. Pollak, North-Holland, Amsterdam 1985, p.1.

\bibitem{wind} S. Wind, M. J. Rooks, V. Chandrasekhar, and
D. E. Prober, Phys. Rev. Lett. {\bf 57}, 633 (1986);
J.J. Lin and N. Giodarno, Phys. Rev. B {\bf 35}, 1071 (1987);
L. Pooke, N. Paquin, M. Pepper, and A. J. Gundlach, Phys. Condens. Matter
{\bf 1}, 3289 (1989); P.M. Etchernach, M.E. Gershenson, H.M. Bozler, A.L.
Bogdanov, and B. Nilsson, Phys. Rev. B {\bf 48} 11516 (1993). 

\bibitem{bergmann} For a review of earlier works, see {\em e. g.}
G. Bergmann, Phys. Rep. {\bf 107} 1 (1984).

\bibitem{mohanty1} P. Mohanty, E. M. Q. Jariwala, and
R. A. Webb, Phys. Rev. Lett. {\bf 78} 3366 (1997); P. Mohanty and
R.A. Webb, Phys. Rev. B {\bf 55}, R13452 (1997).

\bibitem{lin} J. J. Lin and L. Y. Kao, cond-mat/0007417.

\bibitem{doug} D. Natelson, R.J. Willet, K.W. West, and 
L.N. Pfeiffer, cond-mat/0006302.

\bibitem{saclay} A.B. Gougam, F.Pierre, H. Pothier, D. Esteve, and N.O. Birge, 
J. Low Temp.  Phys. {\bf 118}, 447 (2000).


\bibitem{golubev} D. S. Golubev and A. D. Zaikin, Phys. Rev. Lett. 
{\bf 81} 1074 (1998).

\bibitem{aleiner} B. L. Altshuler, M. E. Gershenson, and I. L. Aleiner,
Physica (Amsterdam), {\bf 3E} 58 (1998);  I. L. Aleiner, B. L. Altshuler, and
M. E. Gershenson, (comment on Ref.\cite{golubev}), Phys. Rev. Lett., 
{\bf 82} 3190 (1999); See also the reply of Golubev and Zaikin,
 {\em ibid.}, {\bf 82} 3191 (1999).

\bibitem{persistent-current} P. Mohanty, Ann. Phys (Leipzig) {\bf 8}, 549 (1999);
V.E. Kravtsov and B.L. Altshuler, Phys. Rev. Lett. {\bf 84}, 3394 (2000); P. Schwab,
cons-mat/0005525; P. Cedraschi, V.V. Ponomarenko, and M. Buttiker, Phys. Rev. Lett.
{\bf 84}, 346 (2000).


\bibitem{mohanty2} P. Mohanty, Physica B {\bf 280}, 446 (2000).

\bibitem{kapitulnik}A. Kapitulnik, N. Mason, S.A. Kivelson, and S. Chakravarty, 
cond-mat/0009201.

\bibitem{meir}Y. Meir, Phys. Rev. Lett. {\bf 83}, 3506 (1999).

\bibitem{mohanty3} P. Mohanty, in the Proceedings of the NATO Advanced
Workshop on Size-Dependent Magnetic Scattering, Kluwer Academic Publishers (2000).

\bibitem{ova} A. Vaknin, A. Frydman, and Z. Ovadyahu, Phys. Rev. B {\bf 61}, 13037 (2000).


\bibitem{imry} Y. Imry, H. Fukuyama, and P. Schwab, Europhys. Lett.,
{\bf 47} (5), 608 (1999).


\bibitem{zawadowski} A. Zawadowski, Jan von Delft, and D. C. Ralph,
Phys. Rev. Lett., {\bf 83}, 2632 (1999).

\bibitem{anderson} P. W. Anderson, C. M. Varma, and B. I. Halperin,
Philos. Mag, {\bf 25}, 1 (1972).

\bibitem{phillips0} W. Phillips, J. Low Temp. Phys. {\bf 7}, 351 (1972). 


\bibitem{phillips} For a  review, see {\em e. g.}, 
W. A. Phillips, Rep. Prog. Phys. {\bf 50}, 1657 (1987), and for a more 
recent review, see  also 
{\em Tunneling systems in amorphous and
crystalline solids}, edited by Pablo Esquinazi (Springer, 1998). 

\bibitem{huibers} A. G. Huibers, J.A. Folk, S.R. Patel, C.M. Marcus, 
C.I. Duruoz, and J.S. Harris Jr.., Phys. Rev. Lett. {\bf 81} 200 (1998),
{\em ibid}. {\bf 83} 5090 (1999).

\bibitem{bird} J. P. Bird, K. Ishibashi, D. K. Ferry, Y. Ochiai, 
Y. Aoyagi, and T. Sugano, Phys. Rev. B {\bf 51}, 18037 (1995).

\bibitem{clarke} R.M. Clarke, I. H. Chan, C. M. Marcus, 
C. I. Duru\"oz, J. S. Harris, Jr., K. Campman, and
A. C. Gossard, Phys. Rev. B {\bf 52} 2656 (1995).


\bibitem{buks}
 E. Buks, R. Schuster, M. Heiblum, D. Mahalu, and
V. Umansky, Nature, {\bf 391}, 871 (1998).


\bibitem{zimmerli} G. Zimmerli, T. M. Eiles, R. L.Kautz, and
John M. Martinis, Appl. Phys. Lett. {\bf 61} (2), 237 (1992).

\bibitem{furlan} M. Furlan, T. Heinzel, B. Jeanneret, S. V. Lotkhov, and
K. Ensslin, Europhys. Lett., {\bf 49} (3), 369 (2000).

\bibitem{grupp} D. E. Grupp, T. Zhang, G. J. Dolan, Ned S. Wingreen,
cond-mat/9906023, and references therein.

\bibitem{baranger} H.U. Baranger and P. A. Mello, Phys. Rev. B {\bf 51},
4703 (1995).

\bibitem{brouwer} P. W. Brouwer and C. W. J. Beenakker, Phys. Rev. B
{\bf 55}, 4695 (1997).

\bibitem{stern} A. Stern, Y. Aharonov, and Y. Imry, Phys. Rev. A {\bf 41},
 3436 (1990); A. Stern, Y. Aharonov, and Y. Imry, 
in {\em Quantum Coherence in Mesoscopic Systems} edited by
B. Kramer, Plenum Press, New York, 1991. 

\bibitem{sivan2} U. Sivan,  Y. Imry, and A. G. Aronov, 
Europhys. Lett., {\bf 28}, 115 (1994).

\bibitem{comment1} A. Zaikin and Yu. Nazarov (private communication).

\bibitem{yacoby}
 A. Yacoby, M. Heiblum,
D. Mahalu, and H. Shtrikman, Phys. Rev. Lett., {\bf 74} 4047 (1995);
R. Schuster, E. Buks, M. Heiblum, D. Mahalu, V. Umansky, and
H. Shtrikman, Nature {\bf 385}, 417 (1997).

\bibitem{levinson} Y. Levinson, Europhys. Lett., {\bf 39}, 
 299 (1997).

\bibitem{aleiner3} I. L. Aleiner, N. S. Wingreen, and Y. Meir, Phys. Rev.
Lett. {\bf 79}, 3740 (1997).

\bibitem{buttiker} M. B\"uttiker and A. M. Martin, Phys. Rev. B {\bf 61},
2737 (2000).

\bibitem{sivan1} U. Sivan,  F.P. Milliken, K. Milkove, 
S. Rishton, Y. Lee, J. M. Hong, V. Boegli, D. Kern, 
and M. Defranza, Europhys. Lett. {\bf 25}, 605 (1994).


\bibitem{stern2} A. Stern, Y. Aharonov, and Y. Imry,
{\em Dephasing of Interference by a Back Reacting Environment} ( preprint ).

\bibitem{chakravarty} S. Chakravarty and A. Schmid, Phys. Rep. {\bf 140},
193 (1986).

\bibitem{cohen} 
D. Cohen and Y. Imry, Phys. Rev. B {\bf 59}, 11143 (1999).

\bibitem{black} see {\em e.g.}, J. L. Black, {\em Glassy Metals I}, 
edited by H.-J. G\"untherodt, H. Beck, Springer, Berlin (1981).

\bibitem{feynman} R.P. Feynman and A.R. Hibbs, 
{\em  Quantum Mechanics and Path Integrals} (McGraw-Hill, New York 1965).

\bibitem{schulman} L.S. Schulman, {\em Techniques and Applications
of Path Integration} (Wiely-Interscience, New York, 1981). 

\bibitem{wurger} A. W\"urger, {\it From coherent tunneling to relaxatiob: dissipative 
quantum dynamics of interacting defects}, Springer tracts in modern physics,
Volume 135 (Springer, 1997).

\bibitem{kleiman} R.N. Kleiman, G. Agnolet, and D.J. Bishop, Phys. Rev. Lett. {\bf 59},
2079 (1987).


\bibitem{phillips_si} W. A. Phillips, Phys. Rev. Lett. {\bf 61}, 2632 (1988).

\bibitem{keyes} R.W. Keyes, Phys. Rev. Lett. {\bf 62}, 1324 (1989).


\bibitem{parpia} R.E. Mihailovich and J.M. Parpia, Phys. Rev. Lett. {\bf 68},
3052 (1992).

\bibitem{darrell} P. Mohanty, D.A. Harrington, K.L. Ekinci, Y.T. Yang, M.J. Murphy, and
M.L. Roukes, to be published; D.A. Harrington, P. Mohanty, and M.L. Roukes, 
Physica B {\bf 284-288}, 2145 (2000).

\bibitem{galperin} I. L. Aleiner, B. L. Altshuler, and Y. M. Galperin,
cond-mat/0010228.



\end{references}
\end{document}